# Deep learning-based classification of high intensity light patterns in photorefractive crystals

Marija Ivanović[1], Ana Mančić[2], Carla Hermann-Avigliano[3], Ljupčo Hadžievski[1] and Aleksandra Maluckov[1]

[1] Vinča Institute of Nuclear Sciences, University of Belgrade, Serbia
[2] Faculty of Sciences and Mathematics, University of Niš, Serbia
[3] Deptamento de Física and Millennium Institute for research in Optics, Facultad de Ciencias Fisicas y Matematicas, Universidad de Chile, Chile

E-mail: sandram@vin.bg.ac.rs

**Abstract**

In this paper, we establish a new scheme for identification and classification of high intensity events generated by the propagation of light through a photorefractive SBN crystal. Among these events, which are the inevitable consequence of the development of modulation instability, are speckling and soliton-like patterns. The usual classifiers, developed on statistical measures, such as the significant intensity, often provide only a partial characterization of these events. Here, we try to overcome this deficiency by implementing the convolution neural network method to relate experimental data of light intensity distribution and corresponding numerical outputs with different high intensity regimes. The train and test sets are formed of experimentally obtained intensity profiles at the crystal output facet and corresponding numerical profiles. The accuracy of detection of speckles reaches maximum value of 100%, while the accuracy of solitons and caustic detection is above 97%. These performances are promising for the creation of neural network based routines for prediction of extreme events in wave media.

## 1. Introduction

Extreme events (EEs) or rogue waves (RWs) are high amplitude or intensity events that appear from nowhere and disappear without a trace. They are characterized with a low probability of occurrence [1]. Originally, the term rogue wave referred to isolated gigantic waves appearing suddenly on the surface of the ocean generating huge damage on ships [2]. These phenomena challenge researchers in diverse fields of natural and social sciences, particularly regarding their generation and predictability [3, 4, 5]. EEs are also studied in a variety of optical systems [6, 7, 8, 9]. They are usually associated with the devastation of the information transport, the coherent energy exchange and modification of the system response to the external conditions. In the reference [10] we explored the appearance of RWs in a SBN photorefractive crystal (PRC) at room temperature. We have found that the synergy of light and PRC within the parameter region above the modulation instability (MI) threshold manifest in the formation of a variety of output light intensity patterns. Bellow the MI threshold, the randomly fluctuating intensity landscape was observed. Managing the input light power and properties of the external applied voltage, deep scarce patterns over the intensity landscape were induced by the electro-optic effect.

The regime of clustering of light patterns into high intensity spots (solitons) or speckling events types were also observed.

The idea to challenge the deep learning to extreme events dates from a few years ago when some of us joined the heart arrhythmia studies [11], as well as, from huge implementation of the deep learning in the study of complex phenomena in physics [12, 13]. The missing stimulus to start realizing this idea was recently published paper in Nature Communications [14], where researchers performed a machine learning based analysis of EEs in an optical fibre. It was shown that neural network could be trained to correlate the spectral and temporal properties of the system and then applied to obtain the probability distribution of the highest temporal peaks in the instability field.

Following that line, our goal was to implement the neural network method to experimental data taken in ref [10] which indicated inevitability of the RWs creation in the photorefractive crystals. We use the convolution neural network (CNN) [15] to relate both 2D experimental images of light intensity distribution and corresponding outputs from the numerical simulations within different high intensity and RWs regimes. The intensity profiles are categorized in four classes: regime without high amplitude events (linear dispersion of light), caustic-like regime (clustering of the light intensity), speckles regime (fragmentation of light into the array of large intensity narrow peaks) and soliton regime (self-trapping of light into a few persistent localized structures).

The idea is to test the CNN for regime recognition and then, by following numerically the light throughout the whole journey in the crystal, we establish a deep learning model, which can provide the prediction of the appearance of different types of high intensity events.

The paper is organized in the following manner. After the Introduction, in Section 2, we present experimental and numerical data used to train our CNN. The details of the network architecture and network training are given in Section 3, followed by evaluation results regarding classification performances (Section 4). Obtained results are discussed in Section 5. Final remarks are listed in the concluding section.

## 2. Experimental setup and numerical model

Photorefractive crystals host different nonlinear phenomena [16, 17]. At the core of these phenomena is the liberation of charge carriers from traps in the crystal by the absorption of photons, the subsequent redistribution of charges due to the internal and external electrical fields, and the modification of the refractive index profile via the electro-optic effect responsible for the self-trapping of the propagating light. Since there is always a limit on the number of carriers, nonlinearities supporting creation of localized solitary modes in photorefractive media are saturable [18, 19].

Photorefractive materials are characterized by an extremely slow response to the changes of the light field in time. A feature that distinguishes PRC from other materials used for optical data recording is their ability to record and update the input information continuously. In order to be able to exploit the properties of PRC in the most efficient way, it is important to understand the influence of high amplitude events on the dynamics of these systems.

The details of the experiment are presented in our previous paper [10]. Briefly, the experiment is performed by injecting Gaussian light beams on a 0.005 % $CeO_2$ doped SBN:75 PRC, using the setup sketched in Fig. 1(a) in [10]. The crystal sample had a transversal area of 5x2 $mm^2$ and a length of 10 mm along which the light propagates. The nonlinear response of the crystal is controlled indirectly by an external voltage applied in the crystal vertical axis, along which the input laser beam of wavelength 532 nm and power 10 μW is polarized. In order to observe the optical patterns at the output facet of the crystal, we used a beam profiler CCD camera installed on a translational stage at the end of the setup. This way, we measure the intensities of the output signal by taking images at the crystal output facet, on the typical scale in the interval 0-255 levels. Here, the grayscale images are processed so that 255 gray level represents the highest intensity.

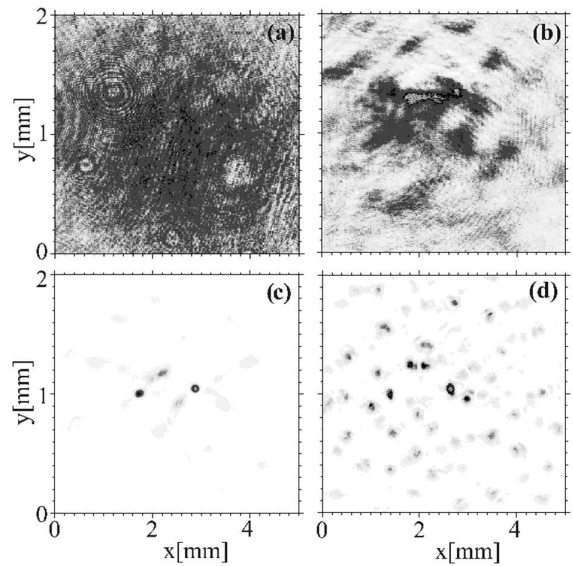

**Figure 1**. Output intensity profiles for 4 regimes obtained experimentally: (a) a dispersion-like, (b) a caustic-like, (c) a soliton-like and (d) a speckling. White and black color corresponds to the lowest and highest intensity, respectively

The experiment shows a rich dynamics of light depending on the input laser beam intensity and shape, as well as on the external voltage parameters such as the actual applied voltage and the speed of change. By managing a small increase rate of the applied voltage, the light spreads smoothly over the crystal that results in the formation of coarse-grained intensity spots. Depending on the input laser power, the crosstalk between spots either smooths the distribution of energy, or it forces the creation of large-intensity structures. The steady state regime is reached by changing the crystal exposure time. Typical experimental output profiles, for a slow variation of the external voltage, are shown in Fig. 1. We distinguished four main output intensity profiles: a dispersion like [Fig.1(a)], a caustic-like [Fig.1(b)], a soliton-like [Fig.1(c)] and a speckling [Fig.1(d)].

From the theoretical point of view, the light propagation through the SBN crystal is modelled by the effective two-dimensional partial differential Schrödinger equation with local saturable nonlinear term, which, in dimensionless form, can be presented as [20]:

$$i\frac{\partial}{\partial z}\psi(x,y,z) + \beta\left(\frac{\partial^2}{\partial x^2} + \frac{\partial^2}{\partial y^2}\right)\psi(x,y,z) - g\frac{\psi(x,y,z)}{1+|\psi(x,y,z)|^2} = 0. \quad (1)$$

$\Psi(x,y,z)$ corresponds to the envelope of the electric field, $x$ and $y$ are transverse crystal sample lengths, and $z$ the propagation coordinate. The second term in Eq. (1) models the light dispersion while the last represents the saturable nonlinear term [21]. The particularity of numerical model is that the only parameter describing the nonlinear crystal response is $g$ and it is a free parameter in the study. Depending on the initial conditions (shape, width and intensity of the laser beam, and the external voltage, here related to $g$) the propagating light can experience different regimes, corresponding to those identified at the output crystal facet in the experiment. Propagation length in the numerical setup is neither fixed nor limited, while in the experiment it is always determined by the length of the crystal.

We showed experimentally and numerically [10] that, in spite of their rarity, RWs are inevitable in our system for input intensities of light above the MI threshold [4]. Moreover, we identified a caustic-like regime which could be a nucleus for high intensity events. Basically, they can be classified as solitary-like waves, characterized by a few persistent huge intensity structures, and speckling regime with very sharp, intensive and randomly distributed peaks. The speckles are usually treated as RWs [see Fig. 1(d)] following the criteria based on the significant intensity ($I_s$) which is defined as the average intensity of one third of the highest intensity waves in the system [22]. All events with intensity higher than $2I_s$ are considered as RWs.

## 3. CNN architecture and training

Numerical calculations offer us a good tool to check the experimental results and to go slightly beyond them. However, standard statistical methods and measures are always related to the determination of the RW threshold by following certain criterion based, more or less, on the observation. Hence, this criterion is approximate and not unique. On the other hand, the deep learning offers a tool for going beyond these limits. The price that has to be paid is our 'passive' role in the 'measurement/detection' procedure and the strong dependence on the input data sets' statistics. By preparing data that are representative enough and well balanced, we can choose the optimal neural network architecture, as well as read and interpret the decision results. But, the details of the inner neural network actions stay hidden.

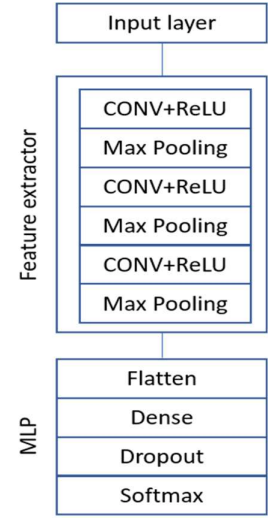

**Figure 2**. Network architecture. Details of MLP are given in ref. [23].

We use a 3-stage feature extractor along with a fully connected multi-layer perceptron (MLP) for recognition of different regimes (Fig. 2). The learning procedure of each feature extraction stage is based on the convolution operation (CONV), which is one of the fundamental operations in the CNNs [15]. Its role is to detect local relations of the previously learned features. The result of convolution operation is then passed through a non-linear activation (the rectified linear unit, ReLU), which, in general, allows the network to learn more complex structures [24]. The final step in each feature extraction stage is max-pooling, which has a role of emphasizing the most significant features. With the three stages of feature extractor we are able to learn the higher-level features by composing the lower ones, learned in the earlier layers, following the general rule that the higher number of the convolutional layers leads to extraction of more complex features. After the high-level features are learned, the classification is obtained by utilizing two-layer perceptron, with a softmax layer as output layer. To improve the generalization capability, dropout is applied between two layers of MLP during the training procedure only [25]. The details of the network architecture are given in Table 1.

**Table 1.** The details of the network architecture; b denotes mini-batch size.

| Layer type | Output shape | # of parameters | Kernel size | Stride/ dropout rate | Activation |
|---|---|---|---|---|---|
| **Input** | (b,512,512,1) | 0 | - | - | - |
| **CONV** | (b,508,508,32) | 832 | 5x5 | 1 | ReLU |
| **MaxPolling** | (b,127,127,32) | 0 | 4x4 | 4 | - |
| **CONV** | (b,123,123,64) | 51264 | 5x5 | 1 | ReLU |
| **MaxPolling** | (b,30,30,64) | 0 | 4x4 | 4 | - |
| **CONV** | (b,26,26,64) | 102464 | 5x5 | 1 | ReLU |
| **MaxPolling** | (b,13,13,64) | 0 | 2x2 | 2 | - |
| **Flatten** | (b,10816) | 0 | - | - | - |
| **Dense** | (b,1024) | 11076608 | - | - | ReLU |
| **Dropout** | (b,1024) | 0 | - | 0.4 | - |
| **Dense** | (b,4) | 4100 | - | - | softmax |

In this study, we prepare the sample set of 1041 experimental output intensity profiles and 969 numerically generated ones. All profiles are divided in four groups according to the results of previous statistical analysis and experience [10]. In order to avoid, as much as possible the uncertainty in this step, several independent supervisions are involved. Each class is more or less equally represented in the sample set, Fig. 3. Both theoretical and experimental datasets are individually split into training and test sets. The 80% of the theoretical data, contained in the training set, is used for choosing an architecture design, tuning the model hyperparameters and evaluating it, with a 10-fold cross-validation. The remaining 20% of the theoretical data is completely withheld for a blindfold testing. The same procedure is repeated for the experimental data alone, and for combined theoretical and experimental data, where the starting architecture is the one selected for theoretical data.

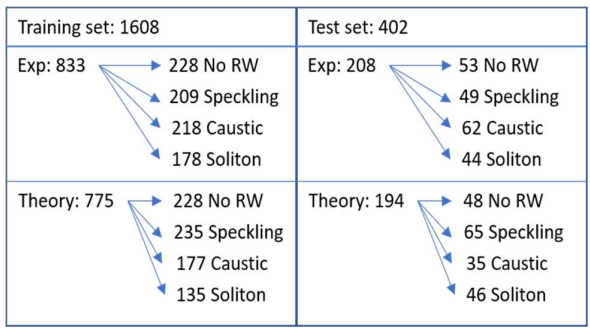

**Figure 3**. Scheme of the dataset content.

The initial weights in the model are obtained with the Xavier normal initializer [26] and are updated by the Adam optimization method [27]. The model is trained by minimizing categorical cross-entropy as a cost function. Training is carried out from scratch in 15 epochs using a mini-batch size of 100 input data.

## 4. Evaluation metrics

Classification performance is evaluated using the overall classification accuracy (Acc), as well as three standard metrics related to each class: class accuracy, sensitivity (Sen) and specificity (Spec). Acc provides a simple way of measuring classifier's overall performance. However, for multi-class classification it does not give enough insight of the performance of each class individually. In order to provide comprehensive assessments of classes' performances, we have included other metrics that are specific to each class. When calculating class performances, the considered class is taken as positives and all other classes as negatives. The metrics are defined as:

$$Acc = \frac{TP+TN}{TP+TN+FP+FN}, \qquad (2)$$

$$Sen \; \frac{TP}{TP+FN}, \qquad (3)$$

$$Spec \; \frac{TN}{TN+FP}, \qquad (4)$$

where TP, TN, FP and FN represent the true positives, true negatives, false positives and false negatives, respectively, with positives corresponding to considered class. In the language of the RWs, in our setup we can distinguish corresponding quantities for each regime (no RWs, caustic-like, soliton-like and speckles). For example, the Acc for speckles is the ratio of the sum of the number of speckling regimes (TP) and all non-speckling regimes (TN: no RW, soliton-like and caustics) correctly predicted by CNN and total number of events (independent on prediction result).

## 5. Results and discussion

In order to evaluate the model and choose an optimal architecture design and model hyperparameters, the averaged performance values over all holdout sets of the 10 fold cross-validation are utilized [28]. The optimal CNN architecture described in Table I is chosen using theoretical training dataset, for which the obtained overall accuracy (Acc) is 97.55 % ± 1.41 %, reported as mean ± standard deviation of 10 fold cross-validation. Afterwards, we used the selected network as a starting architecture for network optimization for experimental data and for combination of both, theoretical and experimental data. It turns out that we do not have to tune any hyperparameter or to change the network architecture since the overall accuracy using the proposed network on experimental and combined training data is 99.76 % ± 0.76 % and 98.69 % ± 1.19 %, respectively. Therefore, the same network shown on Fig. 1 is utilized for evaluating the network performances on the blindfolded test sets of experimental and theoretical data separately as well as combined. Here, we present the model performances evaluated on the blindfolded test sets when the model is trained on the whole training datasets, in the form of the confusion matrix (Fig. 4) and corresponding metric (Table 2.). We pay attention to coordinate training and test sets, meaning that if we want to evaluate the performance on the blindfold experimental data, the whole training set of experimental data has to be used to train the network. Currently, we cannot mix the theoretical and experimental data in sense to use one type for training and the other one for testing since theoretical and experimental data have different distributions. In other words, the experimental data are extracted from the images collected by the CCD camera, therefore each of them is a set of numbers ranging from 0 to 255, i.e. independently on the experimental landscape appearance (noisy background, profile with many narrow high intensity speckles, etc) the full scale of 0-255 is taken into account. On the other hand, the numerical profiles grow from the background, which is supressed, so the relative scaling is not the same as in the experimental images. The preparatory phase of data processing therefore implements the normalization to the maximum in each of the datasets separately.

Regarding the last issue and having in mind that the experimental dataset is a result of real, physical phenomena in the crystal, while the theoretical dataset is a result of the effective mathematical model, we think that the best approach is to develop the CNN analysis on the mix of both experimental and theoretical datasets. In other words, the training and test set should be a balanced mix of experimental and theoretical data. We follow this concept for establishing the preparatory probe for a more developed study of the efficiency of the NN methodology in detecting and predicting extreme events.

By examining the performance metrics given in the Table 2, we come to the following conclusions. The Acc of speckling regime is the highest (100 %) among the accuracies of four regimes. Accordingly, the sensitivity and specificity for that regime are also maximal (100 %). This means that the CNN perfectly detected the speckling profiles, which directly indicate the presence and the relevance of the RWs.

| Theory | | Predicted | | | |
|---|---|---|---|---|---|
| | | noRW | speckling | caustic | soliton |
| TRUE | no RW | 48 | 0 | 0 | 0 |
| | speckling | 0 | 65 | 0 | 0 |
| | caustic | 0 | 0 | 35 | 0 |
| | soliton | 0 | 0 | 12 | 34 |

| Experiment | | Predicted | | | |
|---|---|---|---|---|---|
| | | noRW | speckling | caustic | soliton |
| TRUE | no RW | 52 | 0 | 1 | 0 |
| | speckling | 0 | 49 | 0 | 0 |
| | caustic | 0 | 0 | 62 | 0 |
| | soliton | 0 | 0 | 0 | 44 |

| Theory + Experiment | | Predicted | | | |
|---|---|---|---|---|---|
| | | no RW | speckling | caustic | soliton |
| TRUE | no RW | 100 | 0 | 1 | 0 |
| | speckling | 0 | 114 | 0 | 0 |
| | caustic | 0 | 0 | 97 | 0 |
| | soliton | 0 | 0 | 9 | 81 |

**Figure 4**. Confusion matrices of the test dataset.

When the combination of theoretical and experimental data is used, the Acc of soliton regime of 97.76 % is a consequence of false recognition of 9 profiles as caustic ones. This automatically lowered the sensitivity of the procedure for detection of solitons, while the specificity is 100 % (maximal). The observed discrepancy of Acc and Sen of solitons from 100% is more pronounced for classification based on the theoretical dataset alone, when they reach 93.81 % and 73.91 %, respectively. This is a direct consequence of 12 solitons numerical output regimes falsely classified as caustics ones (Fig. 4). In addition, we directly checked the soliton and caustic numerical profiles by applying the standard statistical measure based on the significant height. Making the strict distinction between these two regimes is a rather tricky endeavour. Roughly, the caustic regime is a transient regime in the numerical calculations. This can be easily demonstrated by not limiting the calculation time to the experimentally proposed value (the fixed length of the crystal). So, the 'problematic' profiles are somewhere on the borderline where solitons start to form as the local increase of intensity, but the final state characterized with persistent solitary structure has not been reached yet.

Table 2. Performance metrics of the test datasets, values of Acc, Sen and Spe are given in %.

| Metrics test set | Theory | Experiment | Theory+experiment |
|---|---|---|---|
| overall Acc | 93.81 | 99.52 | 97.51 |
| Acc no RW | 100.00 | 99.52 | 99.75 |
| Acc speckling | 100.00 | 100.00 | 100.00 |
| Acc caustic | 93.81 | 99.52 | 97.51 |
| Acc soliton | 93.81 | 100.00 | 97.76 |
| Sen no RW | 100.00 | 98.11 | 99.01 |
| Sen speckling | 100.00 | 100.00 | 100.00 |
| Sen caustic | 100.00 | 100.00 | 100.00 |
| Sen soliton | 73.91 | 100.00 | 90.00 |
| Spe no RW | 100.00 | 100.00 | 100.00 |
| Spe speckling | 100.00 | 100.00 | 100.00 |
| Spe caustic | 92.45 | 99.32 | 96.72 |
| Spe soliton | 100.00 | 100.00 | 100.00 |

The accuracy of caustics is 97.51 % due to the false association to caustic profiles of 1 no RWs and 9 soliton profiles. This is also indicated by the specificity of the order of 96.72 %. Again, the numerical data are mostly responsible for that, where Acc=93,81 % and Spe=92.45 % for caustics. There are two reasons that could explain these results. The first one is the same as the one given above for solitons and the second one comes from the fact that caustic phase is a transient one originating from the clustering of the light. However, the extraction metrics are not so affected as in the case of solitons, probably because the appearance of high intensity solitary events is characterized with longer characteristic time.

Finally, the regime with no high intensity events (i.e. no RWs) is characterized by Acc=99.75 %, Sen=99.01 % and Spe=100. It is related to one falsely predicted caustic regime (see Fig. 4). All detected discrepancies are artefacts of the selected parameters for numerical calculations, which are related to the initial idea to model the experimental data by simple theoretical model. However, the mixing of theoretical and experimental datasets in some sense eliminate those discrepancies, which is an interesting and promising result for some future investigations attempting to go towards the predictability of the RWs like events.

Results presented here indicate high performances of the CNN for classification of different high intensity regimes in the propagation of light through a SBN crystal. These offer the opportunity not only to identify the appearance of a rare, high intensity event (RW) but eventually to clarify its properties. Both features could be crucial for the applications.

## 6. Conclusions

In this paper we have done a step ahead towards the implementation of the huge potentials of deep learning methods for the investigation and, hence, for the prediction of the intriguing phenomena of extreme events, or rogue waves. The CNN architecture which consists of the 3-stage feature extractor and a fully connected multi-layer perceptron is applied in order to classify different high intensity profiles formed in the experiment, as a result of light propagation through the SBN crystal, and in the corresponding numerical model. Each feature learning stage incorporates the convolution, ReLU nonlinear activation and max-pooling. Three high intensity profiles: caustic-, soliton- and speckling-like are confronted to the linear dispersion one (i. e. no RWs regime). The network architecture and optimal hyperparameters were selected using 10 fold cross-validation. The model performances are evaluated on the blindfolded test set after the model was trained on the whole training set. When the combination of theoretical and experimental data is considered, the overall accuracy of selecting the soliton and speckling regimes, which can be associated with different types of extreme events is above 97 %. The caustic regime which can be considered as a nucleus for high intensity events is extracted correctly from the other regimes, too: Acc=97.51 %, Spe=96.72 % and Sen=100 %. Satisfying performances of the CNN based detector and classifier of the high intensity events is a stimulating outcome for continuing with the implementation of the deep learning methods in the field of extreme events in different media. We plan to go towards deep learning that will detect, in advance, the system preferences for the formation of high intensity events. This is an important matter to deal with since these events usually have a devastating effect in optical systems.

**Acknowledgements**

We thank our Chilean colleagues for permission to use the experimental results collected in their laboratory during the preparation of the joint publication [10]. We would like to acknowledge the fruitful discussions with R. A. Vicencio and C. Mejía-Cortės. The authors acknowledge the support from the Ministry of Education, Science and Technological Development of Serbia (III 45010), from the Programa ICM Millennium Institute for Research in Optics (MIRO), from U-Inicia VID Universidad de Chile (UI 004/2018), from Comisión Nacional de Investigación Cientifica y Technológica (CONICYT PAI Grant 77180003) and from the European Union's Horizon 2020 Research and Innovation Programme under the Marie Sklodowska-Curie grant agreement No 691051.

**References**

[1] Akhmediev et al. 2016 Roadmap on optical rogue waves and extreme events *J. Opt.* **18** 063001
[2] Onorato M, Osborne A R, Serio M and Bertone S 2001 Freak waves in random seas *Phys. Rev. Lett.* **86** 5831
[3] Mathis A, Froehly L, Toenger S, Dias F, Genty G and Dudley J M 2015 Caustic and Rogue Waves in an Optical Sea *Sci. Rep.* **5** 12822
[4] Akhmediev N and Ankiewicz A 2011 Modulation instability, Fermi-Pasta-Ulam recurrence, rogue waves, nonlinear phase shift, and exact solutions of the Ablowitz-Ladik equation *Phys. Rev. E* **83** 046603
[5] Albevario S, Jentsch V and Kantz H 2006 *Extreme Events in Nature and Society* (Berlin, Springer)
[6] Dudley J M , Genty G and Eggleton B J 2008 Harnessing and control of optical roguewaves in supercontinuum generation *Opt. Express* **16** 3644
[7] Soli D R, Ropers C, Koonath P and Jalali B 2007 Optical rogue waves *Nature* **450** 1054
[8] Marsal N, Caullet V, Wolfersberger D and Sciamanna M 2014 Spatial rogue waves in a photorefractive pattern-forming system *Opt. Lett.* **39** 3690
[9] Suret P, Koussaifi E R, Tikan A, Evain C, Randoux S, Szwaj C and Bielawski S 2016 Single-shot observation of optical rogue wavesin integrable turbulence using time microscopy *Nat. Commun.* **7** 13136
[10] Hermann-Avigliano C, Salinas I A, Rivas D A, Real B, Mančić A, Mejía-Cortės C, Maluckov A and Vincencio R A 2019 Spatial rogue waves in photorefractive SBN crystals *Opt. Lett.* **44** 2807
[11] Ivanovic M, AtanasoskiV, Shvilkin A, Hadzievski Lj and Maluckov A 2019 Deep Learning Approach for Highly Specific Atrial Fibrillation and Flutter Detection based on RR Intervals *41st International Engineering in Medicine and Biology Conference*
[12] Pathak J, Hunt B, Girvan M, Lu Z and Ott E 2018 Model-Free Prediction of Large Spatiotemporally Chaotic Systems from Data: A Reservoir Computing Approach *Phys. Rev. Lett.* **120** 024102
[13] Zahavi T, Dikopoltsev A, Moss D, Haham G I Cohen O, Mannor S and Segev M 2018 Deep learning reconstruction of ultrashort pulses *Optica* **5** 666
[14] Närhi M, Salmela L, Toivonen J, Billet C, Dudley J M and Genty G 2018 Machine learning analysis of extreme events in optical fibre modulation instability *Nat. Commun.* **9** 4923
[15] LeCun Y, Bengio Y and Hinton G 2015 Deep learning *Nature* **521** 436
[16] Günter P, Eichler H J 1987 Introduction to Photorefractive Materials. In: Günter P (eds) Electro-optic and Photorefractive Materials. Springer Proceedings in Physics **18** Springer, Berlin, Heidelberg
[17] Desyatnikov A S, Neshev D N, Kivshar Y S, Sagemerten N, Träger D, Jäagers J, Denz C and Kartashov Y V 2005 Nonlinear photonic lattices in anisotropic nonlocal self-focusing media *Opt. Lett.* **30** 869
[18] Saffman M, McCarthy G and Królikowski W 2004 Two-dimensional modulational instability in photorefractive media *J. Opt. B: Quantum Semiclass. Opt.* **6** S397
[19] Efremidis N K, Sears S, Christodoulides D N, Fleischer J W and Segev M 2002 Discrete solitons in photorefractive optically induced photonic lattices *Phys. Rev E* **66** 046602
[20] Allio R, Guzmán-Silva D, Cantillano C, Morales-Inostroza L, López-González L, Etcheverry S, Vicencio R A and Armijo J 2015 Photorefractive writing and probing of anisotropic linear and nonlinear lattices *J. Opt.* **17** 025101
[21] Onorato M, Residori S, Bortolozzo U, Montina A and Arecchi F T 2013 Rogue waves and their generating mechanisms in different physical contexts *Phys. Rep.* **528** 47
[22] Kharif C and Pelinovsky E 2003 Physical mechanisms of the rogue wave phenomenon *Eur. J. Mech. B Fluids* **22** 603
[23] https://keras.io/layers/core/
[24] Nair V, Hinton G 2010 Rectified linear units improve restricted Boltzmann machines. 27th International Conference on Machine Learning
[25] Srivastava N, Hinton G, Krizhevsky A, Sutskever I, Salakhutdinov R 2014 Dropout: A Simple Way to Prevent Neural Networks from Overfitting *J. Mach. Learn. Res.* **15** 1929
[26] Kingma D P, Ba J 2014 arXiv: 1412.6980
[27] Glorot X and Bengio Y 2010 Understanding the difficulty of training deep feedforward neural *networks* 13th International conference on artificial intelligence and statistics
[28] Hastie T, Tibshirani R and Friedman J 2008 *The elements of statistical learning* (Springer-Verlag, New York)